\documentclass[rnote]{aa}

\usepackage{graphicx}
\usepackage{txfonts}
\usepackage{natbib}
\bibpunct{(}{)}{;}{a}{}{,}

\newcommand{\Teff}{\ensuremath{T_{\rm eff}}}             
\newcommand{\logg}{\ensuremath{\log g}}                  
\newcommand{\vsini}{\ensuremath{v\sin i}}
\newcommand{\Msun}{\ensuremath{\,{\rm M}_\odot}}         
\newcommand{\Rsun}{\ensuremath{\,{\rm R}_\odot}}         

\newcommand{\kms}{\,km\,s$^{-1}$}                        

\begin{document} 

\title{Constrained fitting of disentangled binary star spectra: \\ application to V615\,Per 
      in the open cluster h\,Persei}
\titlerunning{Genetic fitting of disentangled spectra}

\author{ E. Tamajo \inst{1} \and K. Pavlovski \inst{1,2} \and J. Southworth \inst{2} }
\authorrunning{E.\ Tamajo, K.\ Pavlovski \& J.\ Southworth}

\institute{Department of Physics, University of Zagreb, Bijeni\v{c}ka 32, 10\,000 Zagreb, Croatia
           \and Astrophysics Group, Keele University, Newcastle-under-Lyme, ST5 5BG, UK}


\abstract{Using the technique of spectral disentangling, it is possible to determine the individual spectra 
of the components of a multiple star system from composite spectra observed at a range of orbital phases. 
This method has several advantages: it is unaffected by line blending, does not use template spectra, 
and returns individual component spectra with very high signal-to-noise ratios.}
{The disentangled spectra of a binary star system are very well suited to spectroscopic analysis but 
for one problem: the absolute spectral line depths are unknown because this information is not contained 
in the original spectra (unless there is one taken in eclipse) without making assumptions about 
the spectral characteristics of the component stars. Here we present a method for obtaining the atmospheric
 parameters of the component stars by the constrained fitting of synthetic spectra to observed
 and disentangled spectra.}
{Disentangled spectra are fitted using synthetic spectra and a genetic algorithm in order to determine 
the effective temperatures, surface gravities and relative light contributions of the two stars in
 a binary system. The method is demonstrated on synthetic spectra and then applied to the eclipsing 
binary V615\,Per, a member of the young open cluster NGC\,869 (h\,Persei).}
{The method works well for disentangled spectra with signal-to-noise ratios of 100 or more. For V615\,Per 
we find a normal He abundance but an Mg abundance, which indicates bulk metallicity, a factor of two lower 
than typical for nearby OB stars.}
{}


\maketitle 

\section{Introduction}                                                        \label{sec:intro}

The technique of {\em spectral disentangling} {(\sc spd)} allows the isolation of the individual spectra 
of the component stars of a double-lined spectroscopic binary system from a set of composite spectra 
observed over a range of orbital phases. It was originally formulated in the wavelength domain by 
\citet{SimonSturm94aa} and in the Fourier domain by \citet{Hadrava95aas}. The technique simultaneously 
returns the best-fitting individual spectra and the orbital velocity amplitudes of the two stars. 
A detailed overview of {\sc spd} can be found in \citet{PavlovskiHensberge09xxx}.

Compared to other methods of radial velocity measurement, {\sc spd} has several advantages. Firstly, 
it is independent of template spectra so avoids any systematic errors due to spectral differences between
 the target and template stars. Secondly, it is not affected by the blending of spectral lines of the two 
stars \citep[see][]{Hensberge++00aa,MeClausen07aa}. Thirdly, the resulting disentangled spectra contain 
the combined signal of the input spectra \citep{PavlovskiMe09mn} so have a much higher signal-to-noise (S/N) ratio.

There are two disadvantages of the {\sc spd} approach. The first of these is that the continuum normalisation
 of the input spectra has to be very good in order to avoid low-frequency spurious patterns in the resulting 
disentangled spectra \citep{Hensberge++08aa}. The second is that relative continuum light contributions 
of the two stars cannot be found using {\sc spd} as this information is itself not contained in the observed
 spectra, unless a spectrum has been obtained during an eclipse \citep{Ilijic+04aspc}.

{\sc spd} is well suited to the spectral analysis of stars in binary systems. Each disentangled spectrum 
contains only features due to one star, so can be analysed using standard methods for single stars. 
The high S/N ratios of disentangled spectra are very helpful to this process, but the undetermined continuum 
light ratio between the component stars complicates the spectral analysis. In this work we present a method
 to fit synthetic spectra to disentangled spectra, where the atmospheric parameters of the stars are determined
 simultaneously with the relative light contributions of the stars. A genetic algorithm is used for the 
optimization in order to ensure that the best solution is found in a parameter space which suffers from
 strong degeneracies, in particular between effective temperature (\Teff) and surface gravity (\logg).

An important application of {\sc spd} is the study of detached eclipsing binary star systems (dEBs). 
These represent the primary source of directly-measured masses and radii of stars, and as such are 
cornerstones of stellar physics \citep{Andersen91aarv,Torres++10aarv}. {\sc spd} can be used to measure 
the velocity amplitudes of the stars, which are necessary for the mass and radius measurements, 
simultaneously with obtaining the individual stellar spectra for spectral analysis
 \citep{PavlovskiHensberge05aa, Pavlovski+09mn, PavlovskiMe09mn}. A major advantage of dEBs to this
 process is that the surface gravities of the stars can be obtained to within $\pm$0.01\,dex from 
the mass and radius measurements: these parameters can then be fixed in the spectral analysis and 
thus the degeneracy between \Teff\ and \logg\ avoided \citep{Simon++94aa,Hensberge++00aa}.

In this work we demonstrate the genetic algorithm approach to fitting disentangled spectra on 
the dEB V615\,Persei, a member of the young open cluster h\,Persei. \citet[][hereafter SMS04]{Me++04mn} 
obtained a series of high-resolution spectra of V615\,Per and analysed them with published light curves 
\citep{Krzesinski++99aa} to measure the masses (4.08 and 3.18 \Msun) to accuracies of 2\% and the radii
 (2.29 and 1.90 \Rsun) to 5\%, resulting in surface gravities measured to within 0.05\,dex. 
The \Teff\ values were found to be $15\,000 \pm 500$\,K and $11\,000 \pm 500$\,K. SMS04 found that
 stellar evolutionary models required a subsolar metal abundance ($Z \approx 0.01$) to reproduce 
the measured masses and radii of V615\,Per.

The Perseus Double Cluster comprises h\,Persei (NGC\,869) and $\chi$\,Persei (NGC\,884). It has been
 extensively studied via deep CCD photometry \citep{Keller+01aj, MarcoBernabeu01aa, Slesnick+02apj, 
CapillaFabregat02aa, Currie+10apjs}, from which there is general agreement on its distance (2.3 to 2.4 kpc)
 and age (13--14\,Myr). But these studies assumed a solar chemical composition, and their results may be
 systematically wrong if this assumption is incorrect.

Conflicting results on the chemical composition of the Perseus Double Cluster are present in the literature.
 Detailed abundance analyses based on high-resolution spectra of hot stars \citep{Lennon++88aa, Dufton+90aa}
 have challenged previous findings of low helium abundances 
\citep{Nissen76aa, KlochkovaPanchuk87sval, WolffHeasley85apj}.  \citet{Dufton+90aa} and 
\citet{SmarttRolleston97apj} found an approximately solar metal abundance from high-resolution spectra, 
but this was not supported by \citet{Vrancken+00aa}. 
In this work we attempt to shed additional light on this subject by measuring the helium and metal abundances 
of the stars in the dEB V615\,Per.


\section{Constrained fitting of disentangled spectra using a genetic algorithm}

\begin{figure} \includegraphics[width=90mm,height=55mm]{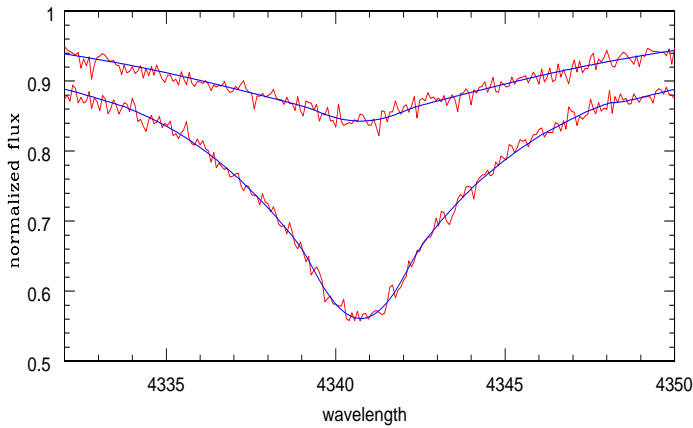} \\
\caption{\label{fig:synHgamma} Synthesized H$\gamma$ line profiles (red
lines) and the best-fitting synthetic spectra (blue lines) for the case
with S/N $=$ 100 in Table\,\ref{tab:synresults1}. The primary star
is below the secondary star on the plot.} \end{figure}

\begin{table*} \begin{center}
\caption{\label{tab:synresults1} Results of fitting synthesized disentangled spectra with {\sc genfitt}
for different S/N values. The output parameters and uncertainties are given, along with the difference
compared to the input parameters and the reduced $\chi^2$ ($\chi^2_\nu$).}
\setlength{\tabcolsep}{6pt}
\begin{tabular}{lrrrrrrrrrrrrc}
\hline \hline
S/N       & \Teff$_1$ (K) & $\Delta$ \Teff$_1$ & $\log g_1$ & $\Delta$ $\log g_1$ & LF$_1$ & $\Delta$ LF$_1$ &
            \Teff$_2$ (K) & $\Delta$ \Teff$_2$ & $\log g_2$ & $\Delta$ $\log g_2$ & LF$_2$ & $\Delta$ LF$_2$ & $\chi^2_\nu$ \\
\hline
$\infty$  & 17\,991  &    $-$9 & 3.495      & $-$0.005 & 0.775      & $-$0.005 & 11\,005  &    $-$5 & 3.995      & $-$0.005 & 0.225      & $+$0.005 & --    \\
          & $\pm$56  &         & $\pm$0.016 &          & $\pm$0.012 &          & $\pm$56  &         & $\pm$0.016 &          & $\pm$0.012 &          &       \\
500       & 17\,896  &  $-$104 & 3.492      & $-$0.008 & 0.771      & $-$0.009 & 10\,896  &  $-$103 & 3.992      & $-$0.008 & 0.229      & $+$0.009 & 1.054 \\
          & $\pm$153 &         & $\pm$0.025 &          & $\pm$0.024 &          & $\pm$157 &         & $\pm$0.023 &          & $\pm$0.021 &          &       \\
200       & 17\,835  &  $-$165 & 3.497      & $-$0.003 & 0.771      & $-$0.009 & 10837    &  $-$164 & 3.987      & $-$0.013 & 0.229      & $+$0.009 & 1.032 \\
          & $\pm$198 &         & $\pm$0.027 &          & $\pm$0.015 &          & $\pm$211 &         & $\pm$0.024 &          & $\pm$0.019 &          &       \\
100       & 17\,753  &  $-$247 & 3.485      & $-$0.015 & 0.769      & $-$0.011 & 10\,748  &  $-$252 & 3.986      & $-$0.015 & 0.231      & $+$0.011 & 0.841 \\
          & $\pm$365 &         & $\pm$0.033 &          & $\pm$0.023 &          & $\pm$358 &         & $\pm$0.036 &          & $\pm$0.021 &          &       \\
50        & 17\,041  &  $-$959 & 3.421      & $-$0.079 & 0.761      & $-$0.019 & 10\,030  &  $-$970 & 3.929      & $-$0.071 & 0.239      & $+$0.019 & 0.774 \\
          & $\pm$441 &         & $\pm$0.035 &          & $\pm$0.081 &          & $\pm$430 &         & $\pm$0.036 &          & $\pm$0.086 &          &       \\
\hline
\end{tabular} \end{center} \tablefoot{The input atmospheric parameters are:
${\Teff}_1 = 18\,000$\,K, ${\logg}_1 = 3.5$, ${\vsini}_1 = 105$\kms, LF$_1 = 0.78$,
${\Teff}_2 = 11\,000$\,K, ${\logg}_2 = 4.0$, ${\vsini}_2 = 95$\kms, LF$_2 = 0.22$}
\end{table*}

\begin{table*} \begin{center}
\caption{\label{tab:synresults2} Same as Table\,\ref{tab:synresults1} but for fits with \logg\ values for the 
two stars fixed.}
\begin{tabular}{lrrrrrrrrc}
\hline \hline
S/N       & \Teff$_1$ (K) & $\Delta$ \Teff$_1$ & LF$_1$ & $\Delta$ LF$_1$ &
            \Teff$_2$ (K) & $\Delta$ \Teff$_2$ & LF$_2$ & $\Delta$ LF$_2$ & $\chi^2_\nu$ \\
\hline
$\infty$  &  17\,921 & $-$79   & 0.776       & $+$0.001 &  10\,921 & $-$79   & 0.224      & $-$0.001 & --    \\
          & $\pm$109 &         & $\pm$0.010  &          & $\pm$115 &         & $\pm$0.014 &          &       \\
500       &  17\,897 & $-$103  & 0.771       & $-$0.004 &  10\,909 & $-$91   & 0.229      & $+$0.009 & 1.065 \\
          & $\pm$160 &         & $\pm$0.022  &          & $\pm$166 &         & $\pm$0.021 &          &       \\
200       &  17\,825 & $-$175  & 0.769       & $-$0.011 &  10\,808 & $-$192  & 0.231      & $+$0.011 & 1.042 \\
          & $\pm$236 &         & $\pm$ 0.018 &          & $\pm$236 &         & $\pm$0.018 &          &       \\
100       &  17\,750 & $-$250  & 0.767       & $-$0.013 &  10\,738 & $-$262  & 0.233      & $+$0.013 & 0.848 \\
          & $\pm$352 &         & $\pm$0.027  &          & $\pm$343 &         & $\pm$0.025 &          &       \\
50        &  17\,056 & $-$944  & 0.756       & $-$0.024 &  10\,030 & $-$970  & 0.244      & $+$0.024 & 0.777 \\
          & $\pm$416 &         & $\pm$0.082  &          & $\pm$422 &         & $\pm$0.085 &          &       \\
\hline
\end{tabular} \end{center} \end{table*}

A computer code has been constructed which fits synthetic spectra to disentangled spectra of a binary system 
in order to determine the atmospheric parameters \Teff, \logg, projected rotational velocities \vsini, 
Doppler shifts, and the light factors. The light factors are an important part of the analysis, and are 
parameterised as LF, the fraction of the total system light produced by one star for the wavelength or 
wavelength interval under consideration. The LFs for the binary components should sum to unity. 
The determination of the atmospheric parameters represents a difficult optimization problem, for which
 we use a genetic algorithm to minimise the $\chi^2$ of the fit to the data 
\citep{Holland75book,Charbonneau96apjs}. Our implementation is called {\sc genfitt} 
({\sc gen}etic {\sc fitt}ing) and in approach is similar to that of \citet{Mokiem+05aa}. Error 
estimates come from the covariance matrix, which is constructed using the Levenberg-Marquardt method.

In order to save computing time we pre-calculate grids of synthetic spectra. {\sc genfitt} linearly 
interpolates between these in \Teff\ and \logg, and then convolves them with a rotational profile using 
the {\sc rotin.f} code of I.\ Hubeny\footnote{\tt http://nova.astro.umd.edu/index.html}. 
The LTE grid covers \Teff\ from 6000 to 15\,000 K and the non-LTE grid covers \Teff\ from 14\,000 
to 35\,000 K. Both grids contain \logg\ values of 2.5 to 5.0 (cgs), and are stepped by 250\,K in 
\Teff\ and 0.1\,dex in \logg.

There can be strong degeneracies between the fitted atmospheric parameters, most notably \Teff\ 
and \logg\ for Balmer line profiles. This degeneracy between \Teff\ and \logg\ can be avoided 
by analysing dEBs, because their surface gravities can be known to within 0.01\,dex from 
measurements of their masses and radii. In many cases the light ratio of the stars in a dEB 
can be obtained from the light curve analysis \citep[e.g.][]{Me++04mn2}, so the LFs can then 
be fixed to known values\footnote{The light ratio can be poorly determined in some dEBs, and in this 
case obtaining a spectroscopic light ratio is an important part of modelling the light curves. 
For an example see \citet{Me++07aa}.}.

In order to test the performance of {\sc genfitt} we synthesized disentangled spectra covering 
the H$\beta$, H$\gamma$ and H$\delta$ lines, using representative atmospheric parameters
and with Gaussian noise added to produce S/N ratios ranging from 25 to infinity. We then 
used {\sc genfitt} to fit all the atmospheric parameters to the synthesized spectra, with 
the only constraint that the two light factors sum to unity. The results are given in 
Table\,\ref{tab:synresults1} and show that for high S/N ratios ($\geqslant$100) the 
{\sc genfitt} results reproduce the input \Teff\ and \logg\ values satisfactorily. For 
lower S/N ratios the inherent degeneracy of these parameters causes them both to be 
underestimated by our method. By contrast, the LFs are reproduced to well within the 
errorbars for all S/N ratios considered. An example fit is shown in Fig.\,\ref{fig:synHgamma}

We obtained a second set of solutions in an `unconstrained mode' where the LFs were not required 
to sum to unity. The results were, as expected, similar to but slightly poorer than the `constrained 
mode'. Finally, a third set of solutions were made with fixed \logg\ values (Table\,\ref{tab:synresults2}), 
as would often be the case when analysing a dEB. We find that the situation is similar to that for 
the first set of solutions. The main limitation on the quality of these results is the degeneracy 
between \Teff\ and \logg: we find a correlation coefficient of 0.98 between these parameters for 
both components. The correlation with the LFs is much weaker, which is why the LF values are 
reliable even for low-S/N spectra.


\section{Application to V615\,Persei}

{\sc genfitt} has already been used for studying the dEBs V380\,Cyg \citep{Pavlovski+09mn} and V621\,Per 
(Southworth et al.\ 2011, in prep.). In both of these cases spectra of a high S/N ratio were available 
and {\sc genfitt} returned excellent results. Here we challenge it with spectra of a much lower S/N ratio.

25 spectra of the dEB V615\,Per were obtained by SMS04, with a reciprocal dispersion of 0.11\,\AA\,pixel$^{-1}$, 
a resolution of 0.2\,\AA\ and an average S/N of $\approx$50. They cover 4220--4500\AA\ so include H$\gamma$, 
a number of helium lines, and the Mg\,II 4481\AA\ doublet and are well distributed through one orbital cycle.
 {\sc spd} was performed in Fourier space using the {\sc fdbinary}\footnote{\tt http://sail.fer.zep/fdbinary/} 
code \citep{Ilijic+04aspc}. 
The disentangled spectra have S/N values of about 160 for the primary star (star\,A) and 80 for the secondary (star\,B).

Since the spectra of V615\,Per cover only a limited wavelength range, our estimate of the \Teff s of the 
component stars is restricted to the H$\gamma$ line. Helium lines are also good \Teff\ indicators, but 
instead we will use these later to obtain the helium abundance of the binary. Because the available light
 curves of V615\,Per are not definitive, the \logg\ values for the two stars are known to modest accuracies
 of 0.059 and 0.050\,dex. We therefore included them as {\sc genfitt} fitted parameters (solution A), 
along with \Teff. We also obtained a solution B for comparison, where the \logg\ values were fixed to 
those found by SMS04. The LFs were constrained to sum to unity. For \vsini\ we adopted $28 \pm 5$ 
and $8 \pm 5$ \kms\ plus an instrumental broadening of 16\kms\ (SMS04). The results are given 
in Table\,\ref{tab:v615genfitt}.

\begin{table*} \centering
\caption{\label{tab:v615genfitt} Results from {\sc genfitt} analysis of the
H$\gamma$ line for V615\,Per, with a comparison to the values found by SMS04.}
\begin{tabular}{l@{\hspace*{30pt}}cc@{\hspace*{20pt}}cc@{\hspace*{20pt}}cc} \hline \hline
Parameter    &\multicolumn{2}{c}{This work (solution A)}&\multicolumn{2}{c}{This work (solution B)}&     \multicolumn{2}{c}{SMS04}         \\
             &       Star A       &    Star B           &       Star A       &    Star B           &      Star A       &    Star B         \\
\hline
\Teff\ (K)   &  $14\,710 \pm 210$ & $11\,520 \pm 290$   &  $14\,920 \pm 190$ & $11\,420 \pm 250$   & $15\,000 \pm 500$ & $11\,000 \pm 500$ \\
\logg\ [cgs] &  $4.302 \pm 0.035$ & $4.361 \pm 0.030$   &  $4.328$ fixed     & $4.381$ fixed       & $4.328 \pm 0.059$ & $4.381 \pm 0.050$ \\
LF           &  $0.676 \pm 0.006$ & $0.324 \pm 0.006$   &  $0.677 \pm 0.004$ & $0.323 \pm 0.004$   & $0.65 \pm 0.03$   & $0.35 \pm 0.03$   \\
\hline \end{tabular} \end{table*}

\begin{figure} \centering
\includegraphics[width=9cm]{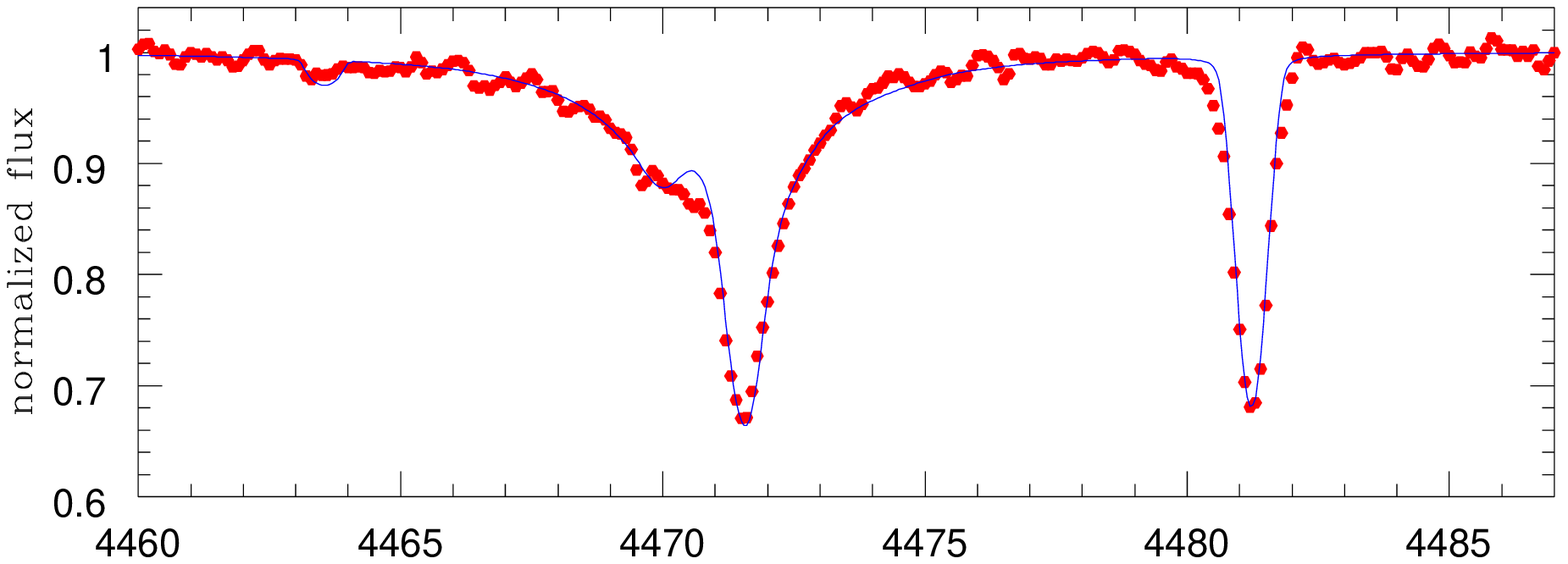}
\includegraphics[width=9cm]{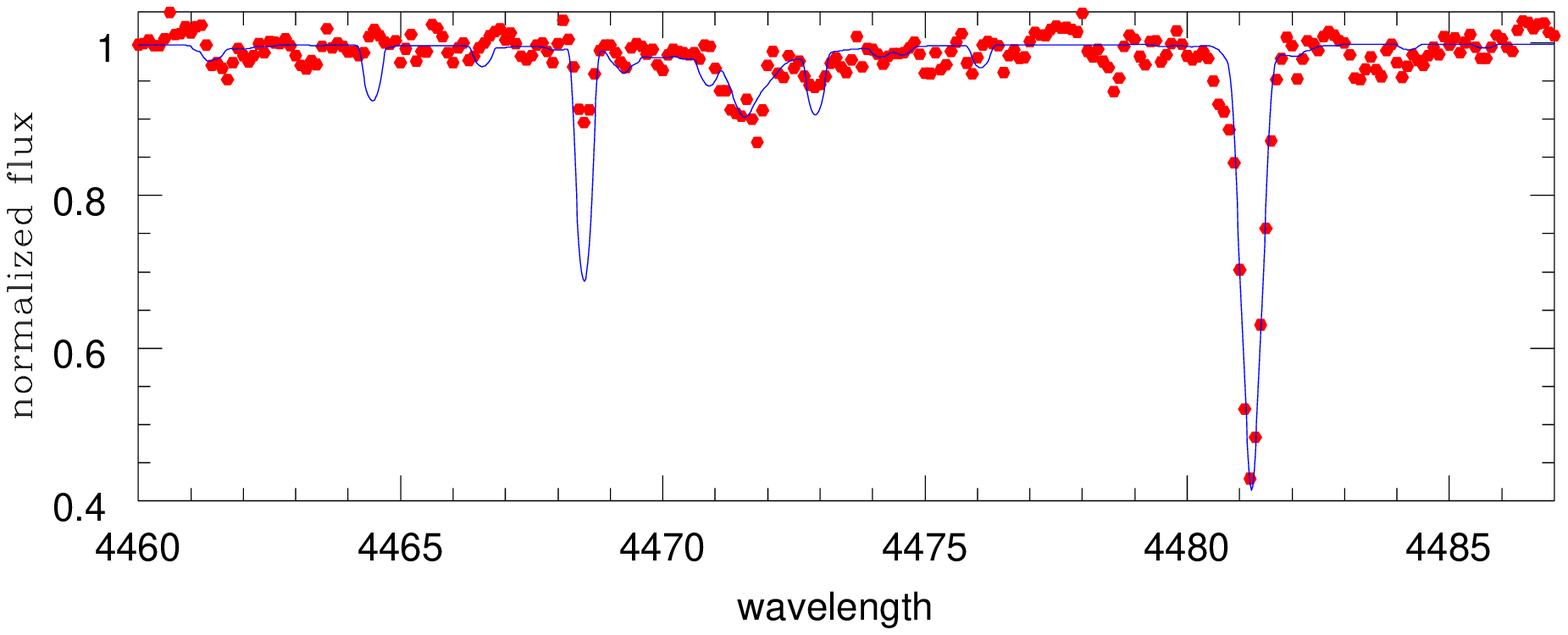} \\
\caption{\label{fig:mg} Comparison between disentangled and renormalised spectra
of star\,A (upper panel) and star\,B (bottom panel) with synthetic spectra
calculated for the He and Mg abundances given in Table\,\ref{tab:v615he}
(blue lines). Only the Mg\,II lines and the \ion{He}{i} line for star\,A are
fitted here. Other lines are shown but not used in our results.} \end{figure}

For the chemical abundance analysis we adopted the atmospheric parameters from solution A. Synthetic 
spectra were calculated using {\sc atlas9} model atmospheres \citep{Kurucz79apjs} and non-LTE theoretical 
line profiles from the {\sc detail} and {\sc surface} codes \citep{Giddings81phd,Butler84phd}. 
A canonical microturbulence velocity of 2\kms\ was adopted \citep{Trundle+07aa} 
as we have too few spectral lines to fit for it.. The helium abundance of star\,A was derived 
via $\chi^2$ minimisation between the observed profiles of He\,I 4388\,\AA\ and 4471\,\AA, 
and profiles calculated for abundances in the range $\epsilon({\rm He}) = 0.06$--$0.15$. 
The results for both lines are given in Table\,\ref{tab:v615he} and correspond to a mean helium 
abundance of $\epsilon({\rm He}) = 0.091\pm007$. This abundance is solar to within the uncertainty 
[$\epsilon_\odot({\rm He}) = 0.089$], so no deviations in helium abundance are detected for V615\,Per\,A. 
The helium lines from star\,B are too weak to be useful (Fig.\,\ref{fig:mg}).

\begin{table} \centering
\caption{\label{tab:v615he} Chemical abundances derived for the components of V615\,Per.}
\begin{tabular}{lccc} \hline \hline
Component & Species & Wavelength (\AA) & Abundance \\
\hline
Star A   & He I  & 4388 & $0.090 \pm 0.005$ \\
Star A   & He I  & 4471 & $0.092 \pm 0.005$ \\
Star A   & He I  & mean & $0.091 \pm 0.007$ \\[3pt]
Star A   & Mg II & 4481 & $7.26  \pm 0.03 $ \\
Star B   & Mg II & 4481 & $7.16  \pm 0.06 $ \\
\hline \end{tabular} \end{table}

Magnesium is an excellent metallicity indicator for B-type stars as it does not participate in the CNO 
process so is unmodified by stellar evolution \citep{Lyubimkov+05mn}. These authors found a mean
 abundance of $\log \epsilon({\rm Mg}) = 7.59\pm0.15$ for nearby B stars with reliable 
microturbulence velocities, in excellent agreement with the solar value of
 $\log \epsilon_{\odot}({\rm Mg}) = 7.55\pm0.02$. Mg\,II 4481\,\AA\ is a prominent 
feature in the spectra of both components of V15\,Per. Theoretical line profiles were calculated 
in non-LTE for star\,A and in LTE for star\,B and the abundances obtained by $\chi^2$ minimisation 
(Fig.\,\ref{fig:mg}). We find a mean abundance of $\log \epsilon({\rm Mg}) = 7.21 \pm 0.07$. 
This is 0.46\,dex lower than the mean value found by \citet{Lyubimkov+05mn} and 0.16\,dex lower
 than that found by \citet{Daflon+03aa} for OB stars in the solar circle. The components of 
V615\,Per have a subsolar Mg abundance and resemble halo B stars more than nearby examples 
\citep{Daflon++04apj}. This implies that the h\,Persei open cluster has a subsolar metal abundance.

\section{Summary}

Spectral disentangling is a method for obtaining the individual spectra of the components of a binary star 
system from composite spectra obtained at a range of orbital phases. A disadvantage of this method is that
 the continuum light ratios of the stars are not found, because this information is not present in the observed
 spectra without making assumptions about the spectral characteristics of the stars. We present the {\sc genfitt}
 program, which uses a genetic algorithm to fit synthetic spectra to the disentangled spectra of both components
 of a binary system simultaneously. It returns the best-fitting atmospheric parameters (\Teff\ and \logg) 
and the light contributions of the two stars. From tests with synthesized spectra we find that {\sc genfitt}
 performs extremely well in determining the light ratio. It also returns reliable \Teff\ and \logg\ values 
in those cases where $S/N$ of the input disentangled spectra is $\geqslant$$100$, which is the usual situation
 for observational studies.

The light contributions of the two stars will normally sum to unity, which provides a useful constraint 
for {\sc genfitt}. Contaminating light from a third star can in principle be found, in cases when the 
light contributions of the two stars in the binary sum to less than unity. Once the light contributions 
of the stars have been found, their disentangled spectra can be renormalised to the correct continuum levels. 
The resulting spectra can then be analysed using standard methods for single stars. If the two stars are 
eclipsing, their surface gravity values may be found to high precision and accuracy from analysis of the 
orbital velocity amplitudes found by spectral disentangling and light curves covering the eclipses.

As a demonstration of the method we applied {\sc genfitt} to spectra of the eclipsing system V615\,Per, 
a member of the h\,Persei open cluster. The metal abundance of this cluster is controversial 
(see Sect.\,\ref{sec:intro}) but important in measuring its distance by the isochrone method. 
The spectra were disentangled and fed into {\sc genfitt}, and an abundance analysis was performed on 
the resulting renormalised spectra. The atmospheric parameters returned by {\sc genfitt} are in good 
agreement with previous work (SMS04) but are more precise. We find a normal solar helium abundance for
 V615\,Per\,A (star\,B is cooler and has only weak helium lines). The magnesium abundances for both
 stars are lower than those found for nearby OB stars, indicating that h\,Persei has a subsolar
 metallicity. This is in agreement with the results of SMS04, based on the complimentary method of 
comparing the masses and radii of the stars to the predictions of theoretical stellar evolutionary models.


\begin{acknowledgements}

KP acknowledges receipt of the Leverhulme Trust Visiting Professorship which enabled him to perform 
this work at Keele University, UK. JS acknowledges funding from STFC in the form of an Advanced Fellowship.

\end{acknowledgements}


\bibliographystyle{aa}

\end{document}